\documentclass[reprint,amsmath,amssymb, aps, prl]{revtex4-2}
\usepackage{graphicx}
\usepackage[colorlinks=true, citecolor=blue, urlcolor=blue]{hyperref}
\usepackage{orcidlink}

\newcommand{\ket}[1]{|#1\rangle}

\newcommand{\innerprod}[2]{\langle{#1}|{#2}\rangle}

\renewcommand{\-}[1]{\mathrm{#1}}

\begin{document}
\preprint{APS/123-QED}
\title{\texorpdfstring{$1/N^2$}{} Precision Interferometry with Collectively Enhanced Atomic Mirror}
\author{Yuan Liu\orcidlink{0009-0006-7349-0903}}
\author{Ke-Mi Xu}
\email{xukemi@bit.edu.cn}
\affiliation{%
MIIT Key Laboratory of Complex-field Intelligent Exploration, School of Optics and Photonics, Beijing Institute of Technology, Beijing 100081, China
}
\author{Hong-Bo Sun}
\email{hbsun@tsinghua.edu.cn}
\author{Linhan Lin}
\email{linlh2019@mail.tsinghua.edu.cn}
\affiliation{%
State Key Laboratory of Precision Measurement Technology and Instruments, Department of Precision Instrument, Tsinghua University, Beijing 100084, China
}
\begin{abstract}
Quantum metrology exploits quantum resources to enhance measurement precision beyond the classical limit. Conventional protocols normally rely on the preparation of delicate quantum states to acquire these resources, posing a major challenge for scaling and robustness. Here we introduce a paradigm that circumvents this requirement with a collectively enhanced quantum mirror (CEAM), i.e., a mesoscopic array of $N$ atoms coupled to a semi-infinite waveguide. When injecting single photons into the waveguide and estimating the CEAM-boundary distance from the reflection phase, a $1/N^2$ precision scaling can be obtained, which surpasses the Heisenberg limit. In this protocol, the quantum resource stems from the cooperative optical response, requiring no entangled state preparation. Our scheme is robust against positional and coupling disorder, offering a practical route to ultra-sensitive quantum metrology in integrated photonic systems.
\end{abstract}

\maketitle
Quantum metrology aims to surpass the classical shot-noise limit by exploiting quantum resources, with the Heisenberg limit ($\propto 1/N$) representing the ultimate scaling achievable with $N$ entangled probes under ideal linear sensing conditions \cite{Giovannetti04S, Giovannetti06Jan, Pezze18RMP, Braun18RMP}. Surpassing this limit--entering the regime of super-Heisenberg metrology--has become a central pursuit, yet it typically demands either nonlinear interactions that consume substantial energy resources \cite{Beltran05PRA, Luis07PRA, Boixo07Feb, Boixo08Jul, Boixo08PRA, Roy08Jun, Napolitano11N} or unusual causal structures that require precise control over complex entangled states \cite{Zhao20May, Yin23NPhys}. These stringent requirements pose significant challenges for robustness and scalability, motivating the search for alternative approaches that can achieve enhanced precision while being resistant against decoherence and perturbations \cite{Ye24May}.

Recently, collective effects in many-body systems have emerged as a promising resource for quantum-enhanced sensing \cite{Qu19PRA, Orioli20PRA, Tudela13Feb, Ostermann13Sep, Reitz22PRXQ, Xu24Sep}. For instance, subradiant states in atomic chains coupled to waveguides exhibit decay rates scaling as $1/N^3$ \cite{Wang25arXiv, Garcia17PRX, Zhang20Dec}, hinting at potential super-Heisenberg sensitivity. However, such schemes restrict the estimated parameter to the interatomic spacing itself, limiting their applicability in integrated photonic circuits.

In this work, we exploit the collective reflection of a mesoscopic atomic array coupled to a semi-infinite waveguide for ultrasensitive distance measurement. Combined with the reflective boundary of the waveguide, this collectively enhanced atomic mirror (CEAM) constitutes an effective optical cavity. The distance $x$ between the CEAM and the boundary, i.e., the cavity length, can then be measured via the phase of reflected single photons. Remarkably, the phase sensitivity scales as $N^2$, leading to a precision $\delta x\propto 1/N^2$ that surpasses the Heisenberg limit. 

The precision improvement can be understood from a unified resource perspective. In standard quantum metrology, the Heisenberg limit is defined with respect to the number of probing particles that directly interact with the parameter. However, from a broader perspective, the entire system—including both the dynamic probes and the static sensor components—can be viewed as a \emph{composite probe}. We fix the dynamic probe to single photons and scale the size $N$ of the static atomic sensor, aiming to provide an optimal environment for the interaction between dynamic probes and the parameters to be measured. This perspective highlights the potential of using structured quantum sensors as a scalable resource for ultra-precision measurements, without the need for delicate quantum state preparation. Our scheme is robust against nonidealities, offering an experimentally feasible approach to ultra-sensitive quantum metrology in waveguide QED systems.

\begin{figure}[htbp]
\centering
\includegraphics[width=8.6cm]{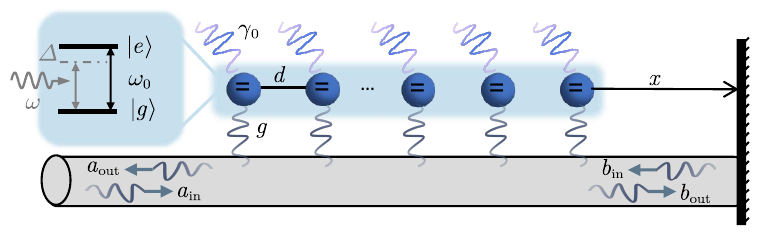}
\caption{Schematic of the system. $N$ atoms (blue dots) coupled to a semi-infinite waveguide terminated by a perfect reflective boundary at distance $x$. The incident single photons ($a_\mathrm{in}$) are reflected from the system.}
\label{figure1}
\end{figure}

\textit{Model--}We consider an array of $N$ identical two-level atoms with transition frequency $\omega_0$ between the ground state $\ket{g}$ and excited state $\ket{e}$; the $j$th atom is described by Pauli operators $\sigma_j^{\pm,z}$. The atom array are coupled to a one-dimensional waveguide with identical strength $V$ for all atoms and placed at a distance $x$ from a perfect reflecting boundary (see Fig.~\ref{figure1}). Single photons with frequency $\omega$ (detuned by $\Delta=\omega_0-\omega$ with respect to the atoms) are incident and then reflected. The atoms are equally spaced by a distance $d=\lambda=2\pi v_\-g/\omega$, where $v_\-g$ is the group velocity of incident photons in the waveguide. The atoms may also decay into non-guided mode or decay nonradiatively, with total rate $\gamma'$ besides coupling to the waveguide \cite{Mahmoodian18Oct, Sheremet23RMP}. For simplicity, we set $\gamma'=0$ in the  theoretical analysis, and includes the influence of nonzero $\gamma'$ in numerical simulations. 

Before analyzing the complete system, we first consider the coupling and scattering properties of the atomic array itself, decoupled from the boundary. In the Markovian and rotating-wave approximations, the array of $N$ identical two-level atoms behaves as a collective quantum object.  The Hamiltonian in a frame rotating at the incident photon frequency $\omega$ is given by \cite{Roy17RMP, Liu25PhotoniX}
\begin{equation}
  H=\sum_{i=1}^{N}\Delta\sigma_i^+\sigma_i^-+\sum_{i=1}^Ng\left[\sigma_i^+(a_\-R\-e^{\-ikx_i}+a_\-L\-e^{-\-ikx_i})+\-{H.c.}\right],
\end{equation}
where the subscript ``R'' and ``L'' denote right- and left-propagating modes, respectively. Due to the equal-distant spacing $d=\lambda$, the atoms interact cooperatively to the guided photon, i.e.,
\begin{equation}
  H=\sum_{i=1}^{N}\Delta\sigma_i^+\sigma_i^-+g_N\left[S^+(a_\-R+a_\-L)+\-{H.c.}\right],
\end{equation}
where we have defined the symmetric collective atomic operator $S^{\pm,z}=\sum_{j=1}^{N}\sigma_j^{\pm,z}/\sqrt{N}$ and coupling strength $g_N=\sqrt{N}g$. In the symmetric single-excitation subspace (considering single photon or weak coherent-light incidence), $S^{\pm,z}$ obey the same commutation relation as $\sigma_j^{\pm,z}$. Consequently the dynamics of the atomic array is equivalent to that of a single two-level system \cite{Houdre96PRA} with an enhanced decay rate into the waveguide $\Gamma=N\gamma$, where $\gamma=2\pi g^2/v_{\mathrm{g}}$ is the single-atom decay rate. This collective enhancement arises from constructive interference of the scattered fields and is responsible for the array's high reflectivity.

\textit{Reflection Coefficients--}The reflection and transmission coefficients of a single atom, and consequently of the CEAM can be derived using standard input–output theory. For a single photon incident from the left, we obtain the reflection and transmission coefficients for the CEAM alone \cite{Sheremet23RMP, Roy17RMP},
\begin{equation}
r=-\frac{\Gamma/2}{\-i\Delta+\Gamma/2}, \qquad
t=\frac{\-i\Delta}{\-i\Delta+\Gamma/2},
\label{eq-rt0}
\end{equation}
with $|r|^2 + |t|^2 = 1$ and $t = 1 + r$ due to energy conservation and time-reversal symmetry. Notably, for resonant excitation ($\Delta = 0$), the CEAM becomes perfectly reflective ($r=-1$ and $t=0$, independent of $N$); otherwise the reflective coefficient $r$ increases and approaches identity when $N\to\infty$. These coefficients serve as the building blocks for analyzing the full system including the reflective boundary.

The total reflection coefficient $R$ of the entire system, i.e., of a cavity with length $X=x+(N-1)d$ formed by CEAM plus boundary, is derived by considering the input-output relation on both sides of the CEAM and the boundary condition at the waveguide end. Let $a_\-{in}$ and $a_\-{out}$ be the incoming and outgoing field amplitudes on the left side of the array, and $b_\-{in}$ and $b_\-{out}$ on the right side. The scattering relations are
\begin{equation}
a_\-{out}=ra_\-{in}+tb_\-{in}, \qquad
b_\-{out}=ta_\-{in}+rb_\-{in}.
\label{eq-scatt}
\end{equation}
The reflective boundary at the waveguide end imposes
\begin{equation}
b_\-{in}=-\-e^{2\-ikx}b_\-{out},
\label{eq-boundary}
\end{equation}
where $k=\omega/v_\-g$ (neglecting small corrections of the group velocity). Substituting Eq.~(\ref{eq-scatt}) into Eq.~(\ref{eq-boundary}) yields
\begin{equation}
b_\-{in}=-\frac{t\-e^{2\-ikx}}{1+r\-e^{2\-ikx}}a_\-{in}.
\label{eq-bin}
\end{equation}
Inserting Eq.~(\ref{eq-bin}) into the expression for $a_\-{out}$ gives the total reflection coefficient:
\begin{equation}
R=\frac{a_\-{out}}{a_\-{in}}=r-\frac{t^2\-e^{2\-ikx}}{1+r\-e^{2\-ikx}}.
\label{eq-R}
\end{equation}

\begin{figure}[htbp]
\centering
\includegraphics[width=8.6cm]{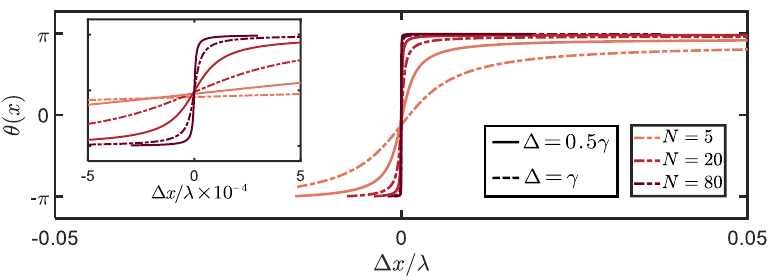}
\caption{Phase sensitivity of the system under different number of atoms and different atom-photon frequency detunings. The inset shows the same data on an expanded $x$‑axis scale.}
\label{figure2}
\end{figure}

\textit{Phase Sensitivity--}The phase of $R$, denoted by $\theta=\arg(R)$, is sensitive to the cavity length $X$. For the ideal case of no deviations in atomic coordinates, the sensitivity is quantified by the derivative $\partial\theta/\partial X=\partial\theta/\partial x$, which can be expressed as
\begin{equation}
\frac{\partial\theta}{\partial x}=\Im\left(\frac{1}{R}\frac{\partial R}{\partial x}\right).
\end{equation}
From Eq.~(\ref{eq-R}), with $z=\-e^{2\-ikx}$, we compute
\begin{equation}
\frac{\partial R}{\partial x}=-\frac{2\-ikzt^2}{(1+zr)^2}.
\label{eq-dRdx}
\end{equation}

To maximize $|\partial\theta/\partial x|$, we choose the working point such that $|1+zr|$ is minimized. This occurs when $zr$ is a negative real number close to $-1$. Setting $r=|r|\-e^{\-i\theta_r}$, we choose $z=\-e^{\-i(\pi-\theta_r)}$ so that
\begin{equation}
zr = -|r|. \label{eq-opt}
\end{equation}
Then $1+zr=1-|r|\equiv\:r$. Since the system is lossless ($|R|=1$), we have $\:r\ll 1$ for large $N$ and fixed $\Delta$. From Eq.~(\ref{eq-rt0}) we obtain
\begin{align}
|r|&=\frac{\Gamma/2}{\sqrt{\Delta^2+(\Gamma/2)^2}}\approx1-\frac{2\Delta^2}{N^2\gamma^2}, \\
t&=\frac{2\-i\Delta}{\Gamma}\cdot\frac{1}{1+2\-i\Delta/\Gamma}\approx\frac{2\-i\Delta}{N\gamma}, 
\end{align}
and consequently 
\begin{equation}
\:r=1-|r|\approx\frac{2\Delta^2}{N^2\gamma^2}.
\label{eq-scaling}
\end{equation}

Now, at the optimal point, using Eqs.~(\ref{eq-dRdx}), (\ref{eq-scaling}), and noting $|z|=1$, we have:
\begin{equation}\label{eq-dRdxorder}
\left|\frac{\partial R}{\partial x}\right|=2k\frac{|t|^2}{\:r^2}=N^2\frac{2k\gamma^2}{\Delta^2}.
\end{equation}
Since $|R|=1$, we also have $\left| \frac{1}{R} \frac{\partial R}{\partial x} \right| \propto N^2$. Consequently, the phase sensitivity scales as
\begin{equation}\label{eq-phase-sensitivity}
\left|\frac{\partial\theta}{\partial x}\right|\propto N^2.
\end{equation}
Such an quadratic scaling on $N$ is visually shown in Fig.~\ref{figure2}, where a drastically enhanced phase gradient can be observed when $N$ is large. Meanwhile, a smaller detuning $\Delta$ also yields an higher sensitivity, which, however, may result in reduced robustness of the system, as will be analyzed later.

\textit{Interferometric Measurement--}To access the phase information $\theta(x)$ imprinted on the reflected photon, we employ a measurement scheme that interferometrically combines the signal field with a reference field, converting the phase shift into a detectable population difference or quadrature signal.

Consider a single-photon pulse prepared in an equal superposition of the signal mode (which interacts with the atomic array and the mirror) and a reference mode. The quantum state of the total system reads
\begin{equation}
  \ket{\Psi}=\ket{\psi_\-{ph}}\ket{g}^{\otimes N},
\end{equation}
where the initial photonic state is
\begin{equation}
\label{eq-initial_state}
\ket{\psi_{\-{ph},0}}=\frac{1}{\sqrt{2}}\bigl(\ket{1_\-s,0_\-r}+\-i\ket{0_\-s,1_\-r}\bigr),
\end{equation}
and $\ket{g}^{\otimes N}$ denotes the ground state of all $N$ atoms. The subscripts s and r denote the signal and reference modes, respectively. After the signal photon is reflected from the atom--mirror system, it acquires a phase shift $\theta(x)$ (while the atomic state remains in $\ket{g}^{\otimes N}$), i.e.,
\begin{equation}
\label{eq-phase_shifted}
\ket{\psi_\-{ph}(x)}=\frac{1}{\sqrt{2}}\bigl(\-e^{\-i\theta(x)}\ket{1_\-s,0_\-r}+\-i\ket{0_\-s,1_\-r}\bigr). 
\end{equation}
The two modes are then combined via a balanced interferometer, and a measurement of the output port populations or quadrature amplitudes yields information about $\theta(x)$. Mathematically, the combination corresponds to a unitary transformation that maps the state to
\begin{equation}
\ket{\psi_\-{out}}=\frac{1}{2}\left[(\-e^{\-i\theta(x)}-1)\ket{1_\-a,0_\-b}+\-i(\-e^{\-i\theta(x)}+1)\ket{0_\-a,1_\-b}\right],
\end{equation}
where the two output modes are labeled a and b. The probability difference between the two output ports is
\begin{equation}
\langle\Sigma_z\rangle=|\innerprod{0_\-a,1_\-b}{\psi_\-{out}}|^2-|\innerprod{1_\-a,0_\-b}{\psi_\-{out}}|^2=\cos\theta(x),
\end{equation}
which oscillates with $\theta(x)$ and therefore allows estimation of the distance $x$.

\textit{Discussion--}The quantum Fisher information quantifies the maximum information about the parameter $x$ that can be extracted from the quantum state $\ket{\Psi}$. For a pure state, the QFI is given by \cite{Pezze18RMP} $F(x)=4\Bigl[\innerprod{\partial_x\Psi}{\partial_x\Psi}-\bigl|\innerprod{\Psi}{\partial_x\Psi}\bigr|^2\Bigr]$. Applying this to the state in Eq.~\eqref{eq-phase_shifted}, we obtain
\begin{equation}
F(x)=\left[\frac{\partial\theta(x)}{\partial x}\right]^2\propto N^4.
\end{equation}
Thus, for our single-photon interferometric scheme, the QFI is simply the square of the derivative of the reflection phase with respect to $x$. 

For any unbiased estimator, the variance of the estimate of $x$ after $\nu$ independent repetitions of the experiment is bounded by the Cram\'er-Rao inequality $\delta x \ge 1/\sqrt{\nu\,F(x)}$, where $\delta x$ denotes the root-mean-square error. Consequently we have
\begin{equation}
\label{eq-precision}
\delta x \ge \frac{1}{\sqrt{\nu}\,|\partial_x\theta(x)|}\propto\frac{1}{\sqrt{\nu}N^2}.
\end{equation}
The lower bound of $\delta x$, which is saturated by the interferometric measurement scheme described above, surpasses the Heisenberg limit ($\propto N^{-1}$) with respect to the number of atoms $N$. In what follows, we give an intuitive explanation on the super-Heisenberg-limited scaling.
 
A standard procedure of quantum phase estimation follows the general scheme \cite{Pezze18RMP, Braun18RMP}: (i) preparing a probe state $\rho$, (ii) encoding on $\rho$ a phase shift $\theta$ that depends on the quantity of interest, ie., $\rho\mapsto\rho_\theta$, (iii) detecting the output state $\rho_\theta$ and (iv) analyzing the measurement outcomes to estimate $\theta$. For a phase generated by $U_\theta=\exp(-\-i\theta H)$ and $\rho_\theta=U_\theta\rho U_\theta^\dagger$, the use of $N$ uncorrelated probes gives rise to a minimum uncertainty $\delta\theta\propto1/\sqrt{N}$ (standard quantum limit); using $N$ entangled probes, on the other hand, permits the Heisenberg-limited scaling $\delta\theta\propto1/N$. For example, considering the Greenberger-Horne-Zeilinger (GHZ) state $\ket{\-{GHZ}}=(\ket{0}^{\otimes N}+\ket{1}^{\otimes N})/\sqrt{2}$ and the Hamiltonian $H=\sigma^z$, the phase encoding yields $U_\theta^{\otimes N}\ket{\-{GHZ}}=(\-e^{-\-iN\theta}\ket{0}^{\otimes N}+\-e^{\-iN\theta}\ket{1}^{\otimes N})/\sqrt{2}$. However, even in the absence of entanglement, one can achieve the $1/N$ scaling by sequential iterations \cite{Tao24Aug, Braun18RMP}; e.g., the single-probe state $\ket{\psi}=(\ket{0}+\ket{1})/\sqrt{2}$ undergoing $N$ cycles of interaction evolves to $U_\theta^N\ket{\psi}=(\-e^{-\-iN\theta}\ket{0}+\-e^{\-iN\theta}\ket{1})/\sqrt{2}$.

Motivated by the insight of sequential iteration, the $1/N^2$ scaling in this work can be intuitively understood by drawing an analogy with a Fabry-P\'erot (FP) cavity. A standard FP cavity consists of two mirrors with amplitude reflectivities $r_1$ and $r_2$. Photons bounce between two mirrors several times before escaping. Effectively, the number of bounces is \cite{Reiserer15RMP}
\begin{equation}
  N_\-b=\frac{\+F}{\pi},
\end{equation}
where the finesse $\+F$ of the cavity \footnote{The finess $\mathcal{F}$ of a conventional FP cavity also characterizes the sharpness of resonances; however, for the atomic cavity here, we have $|r|\simeq1$ for a large range of detune $\Delta$ when $N$ is large, so $\mathcal{F}$ here is only used to relate the amplitude reflection $r$ and the number of photon bounces $N_\mathrm{b}$, regardless of the sharpness of resonances.} is given by
\begin{equation}
  \+F = \frac{\pi r}{1-|r|^2},\quad r=\sqrt{|r_1 r_2|}.
\end{equation}
In the limit of high reflectivity $(|r| \to 1)$, the finesse scales as $\+F \propto (1-|r|^2)^{-1}$. The atomic array exhibits a collective quantum response, where the deviation of the arrays' reflection coefficient from unity scales as $(1-|r|^2)\propto N^{-2}$. Thus the effective ``finesse'' of the cavity with a CEAM, and consequently the number of photon bounces, scales as $N_\-b\propto\+F_{\text{atom}}\propto 1/\:r \propto N^2$, consistent with the phase sensitivity derived in Eq.~\eqref{eq-phase-sensitivity}. 

\begin{figure}[htbp]
\centering
\includegraphics[width=8.6cm]{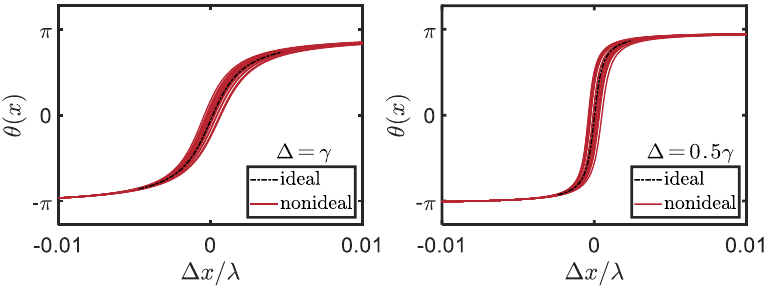}
\caption{Robustness of the phase sensitivity under imperfections for detunings $\Delta=\gamma$ (left) and $\Delta=0.5\gamma$. We considered $N=10$ transmon qubits ($f_0=6$~GHz) coupled to a common waveguide, with $\gamma=2\pi\times100$~MHz, $\gamma'=2\pi\times0.5$~MHz. Imperfections include atom-waveguide coupling variations ($3\sigma=0.03\gamma$), positional disorder ($3\sigma=500$~nm), and transition frequency fluctuations ($3\sigma=100$~MHz), each modeled by a truncated Gaussian distribution within $\pm3\sigma$. For each subplot, 20 random combinations of these imperfections are generated, and the resulting phase sensitivity curves are overlaid. The ideal case (black dashed line) is reserved for comparison.}
\label{figure3}
\end{figure}

\textit{Robustness and Experimental Feasibility--}An ideal quantum sensor should be ultrasensitive to the signal, while remaining insensitive to noise and imperfections \cite{Yang26Feb, Peng24Aug, Zhang23Nov, Yang25NPhotonics}. Here we discuss the robustness against nonidealities in transition frequency, coupling strength and positioning of the atoms. In the Supplemental Materials \footnote{See Supplemental Materials for detailed derivations.}, we have derived the leading-order effects of disorder on the total reflection coefficient $R$ via the transmission matrix approach, i.e.,
\begin{equation}
  \-dR=\sum_{i=1}^N\left[\frac{2\-i\gamma}{\Delta^2}\-d\omega_i+\frac{2\-i}{\Delta}\-d\gamma_i\right]+\sum_{i=1}^{N-1}\frac{N\gamma^2kr}{\Delta^2}\-dx_i,
\end{equation}
where $\-d\omega_i$, $\-d\gamma_i$ and $\-dx_i$ denote, for the $i$-th atom, the deviation in the transition frequency, in the decay rate into the waveguide and the displacement from its ideal position, respectively. The contributions of deviations from different atoms may partially cancel depending on the specific disorder configuration, enhancing the robustness of the scheme. 

For a proof-of-concept experiment, the CEAM can be realized with superconducting transmon qubits coupled to a coplanar waveguide (CPW), as reported in ref.~\cite{Mirhosseini19N}. We consider $N=10$ atoms (transmon qubits), each having a waveguide-mediated decay rate $\gamma/2\pi = 100$~MHz and a parasitic decoherence rate $\gamma'/2\pi \approx 0.5$~MHz, yielding a single-qubit Purcell factor $P_\-{1D}\equiv\gamma/\gamma'=200$. The reflecting boundary is realized by terminating the waveguide with a short circuit at a distance $x$ from the array; this distance can be tuned using an integrated piezoelectric actuator. The transition frequency is $f_0=6$~GHz, corresponding to a wavelength $\lambda_0=5$~cm. In ref. \cite{Mirhosseini19N}, the variation in atom-waveguide coupling strengths between two atoms is characterized as $(\gamma_1-\gamma_2)/(\gamma_1+\gamma_2)\approx 0.03$. For transmon qubits fabricated by electron beam lithography (EBL), the spatial accuracy is within the hundred-nanometer range \cite{Damme24N, Foroozani19QSciTechnol}, which is much smaller than the wavelength $\lambda_0$ even after accounting for the linear scaling of $\-dx_i$ with respect to $N$. Typical static frequency spread of transmon qubits without flux bias tuning is below 100~MHz \cite{Osman23PRR, Zhang22SA, Wang24arxiv}. In Fig.~\ref{figure3}, we take all these nonidealities into consideration, and observe that the sensor remains robust under such a practical setup. Note that although a smaller detuning $\Delta$ yields a higher sensitivity from Eq.~\eqref{eq-dRdxorder}, it also increases the susceptibility to imperfection, suggesting a tradeoff between sensitivity and robustness.

Crucially, all these static offsets only influence the global performance of the sensor (e.g., the sensitivity and the working point) and can be calibrated. For example, the transition frequencies of different transmons can be adjusted to sub MHz level via flux biasing \cite{Burnett19npjQInf, Dhieb25arxiv}, it is only the kHz-level temporal drift during the subsequent experiments that cannot be tracked or compensated in real time. Thus, the proposed sensor maintains a high signal-to-noise ratio even in the presence of moderate imperfections. 

\textit{Conclusion and Perspective--}We have demonstrated that a CEAM coupled to a semi-infinite waveguide can achieve super-Heisenberg scaling in distance measurement without entanglement. The  extremely sensitive reflection phase stems from the constructive interference of the atoms which enhances the optical response of the effective cavity. 

The design principle presented here is not limited to atomic arrays but may be extended to other engineered quantum interfaces, paving the way for robust, scalable quantum sensors. Looking forward, exploring the scenario with multi-photon incident fields could reveal richer collective dynamics and nonlinear optical responses \cite{Mahmoodian18Oct, Iversen21Feb, Schrinski22Sep, Mahmoodian20PRX, Tomm23NPhys}. Moreover, the integration of more sophisticated nonclassical light sources--such as squeezed or entangled photon states \cite{Abbott16Feb, Nehra22S, Dowling08ContempPhys}--with the CEAM may open up new regimes of hybrid quantum sensing, where the combined resource of structured sensor hardware and tailored quantum probes could yield further improvements in sensitivity.

\begin{acknowledgments}
\textit{Acknowledgments--}L.L. acknowledges support from the Fundamental and Interdisciplinary Disciplines Breakthrough Plan of the Ministry of Education of China (grant No. JYB2025XDXM115), the Ministry of Education of China Scientific Research Innovation Capability Support Project for Young Faculty (grant No. SRICSPYF-ZY2025009), the Tsinghua University Initiative Scientific Research Program, the National Key Research and Development Program of China (grant No. 2020YFA0715000), and the National Natural Science Foundation of China (grant No. 62075111); K.-M.X. acknowledges support from the Fundamental and Interdisciplinary Disciplines Breakthrough Plan of the Ministry of Education of China (grant Nos. JYB2025XDXM115 and JYB2025XDXM201) and the Beijing Science and Technology Planning Project (grant No. Z25110100040000); H.-B.S. acknowledges support from the National Natural Science Foundation of China (grant No. 61960206003) and Tsinghua-Foshan Innovation Special Fund (grant No. 2021THFS0102).
\end{acknowledgments}

\vspace{1em}
\textit{Data Availability--}The data that support the findings of this article are not publicly available. The data are available from the authors upon reasonable request.
%
\end{document}